\documentclass[fleqn,10pt,final,twocolumn,twoside,letterpaper,journal,comsoc]{IEEEtran}

\usepackage{graphicx,url}
\usepackage[english]{babel}
\usepackage[utf8]{inputenc}
\usepackage{amsmath}
\usepackage{balance}
\usepackage[margin=2cm]{geometry}
\usepackage{stfloats}
\usepackage{lipsum}
\usepackage{mathtools}
\usepackage{cuted}
\usepackage{graphicx,caption}
\usepackage{caption}
\ifCLASSINFOpdf
\else
\fi

\begin{document}

\title{A Two-Dimensional FFT Precoded Filter Bank Scheme}

\author{\IEEEauthorblockN{Rogério~Pereira~Junior, Carlos~A.~F.~da~Rocha, Bruno~S.~Chang and Didier~Le~Ruyet}

\thanks{Rogério Pereira Junior and Carlos A. F. da Rocha are with the Communications Research Group, Federal University of Santa Catarina, Florianópolis, Brazil, e-mail:rogerio.pereira.junior@posgrad.ufsc.br, carlos.aurelio@ufsc.br; 
Bruno S. Chang is with CPGEI/Electronics Department, Federal University of Technology - Paraná, Curitiba, Brazil, e-mail:bschang@utfpr.edu.br;
Didier Le Ruyet is with CEDRIC, Conservatoire National des Arts et Métiers, Paris, France, e-mail:leruyet@cnam.fr. 
This work was financed in part by the Coordenação de Aperfeiçoamento de Pessoal de Nível Superior – Brasil (CAPES).}
}

\maketitle

{
\selectlanguage{english}
\begin{abstract}
 This work proposes a new precoded filter bank (FB) system via a two-dimensional (2D) fast Fourier transform (2D-FFT). Its structure is similar to Orthogonal Time Frequency Space (OTFS) systems, where the OFDM transmitter is changed to a filter bank multi-carrier (FBMC) one, thus obtaining a lower out-of-band emission. The complex orthogonality of the FBMC transmission is guaranteed by using precoding based on a discrete Fourier transform, which is also used to implement the two-dimensional fast Fourier transform. Through the use of a global transmission matrix, we propose a hybrid receiver for the new system. First, a frequency domain equalization is performed, followed by an interference cancellation on the delay-Doppler domain. The simulation results show that the proposed system obtains an error performance similar to other OTFS systems, and superior performance as compared to other precoded FBMC systems.
\end{abstract}
}

\IEEEpeerreviewmaketitle

\section{Introduction}

Increasing technological development and the emergence of new wireless systems have  a huge impact in human history. Today, there is a broad convergence of processing and messaging technologies.
The fifth generation (5G) of mobile systems is known to provide a huge increase in data rates and a wide variety of application scenarios with respect to the previous generation \cite{alliance20155g}.
With the emergence of these new applications, not only low latencies are sought but also low levels of interference, robustness in high mobility scenarios, in addition to better spectral confinement. Thus, future networks such as sixth generation (6G) mobile systems must be able to expand and meet this demand, introducing new forms of efficient transmission \cite{saad2019vision}. In applications for industry and 4.0 services, such as the Internet of Things (loT), a waveform that has a good spectral location and is also flexible is needed, providing a better and more efficient resource allocation to meet the requirements of different use cases. 
Although it was decided to keep orthogonal frequency division multiplexing (OFDM) as the waveform in 5G systems \cite{parkvall2017nr}, all of these requirements proved difficult to be met by such a technique. Problems with out-of-band (OOB) emissions are recurrent in OFDM, requiring changes in its structure such as filtering to meet the metrics imposed for future wireless systems.

Filter bank multi-carrier (FBMC) is an alternative to combat the high OOB emissions of OFDM \cite{siohan2002analysis}. 
This technique is based on subcarrier filtering and does not use a guard interval, which increases the spectral efficiency. It also improves the spectral location and, consequently, limits the OOB emissions caused by the use of the rectangular window in OFDM \cite{banelli2014modulation}. 
However, the filtering process generates imaginary interference and consequently, the loss of complex orthogonality. Thus, it is necessary to relax the complex orthogonality criterion to the real field, motivating the use of Offset Quadrature Amplitude Modulation (OQAM) \cite{bolcskei2003orthogonal}. 
Through OQAM, the imaginary and the real parts of a complex symbol are transmitted with a shift of half of the symbol period between them. On the receiver side, the data is transported only by the real (or imaginary) component and the intrinsic interference term appears in the imaginary (or real) part. Thus, a transmission of only real symbols at a double rate (to keep the maximum transmission capacity) free from filter interference is generated. However, even though the interference is orthogonal to the data symbols the loss of complex orthogonality leads to problems in the use of multiple input and multiple output (MIMO) techniques \cite{lele2010alamouti}. Moreover, FBMC shares some problems with OFDM, such as high peak to average power ratio (PAPR) and sensitivity to high Doppler spreads \cite{strohmer2003optimal}.
Future wireless systems will operate in highly mobile environments, such as high-speed trains and millimeter wave systems  \cite{boccardi2014five}. 
In this scenario, both OFDM and FBMC present significant interferences, because the channel varies in time and, consequently, can present strong Doppler dispersion \cite{farhang2016ofdm}.

In order to minimize these limitations, precoded versions of OFDM and FBMC systems are presented in the literature. 
Related to OFDM, solutions such as orthogonal time-frequency space modulation (OTFS) \cite{hadani2017orthogonal} were proposed, aiming at applications that require high data rate and mobility. Using a
sympletic finite Fourier transform (SFFT), which is a two-dimensional fast Fourier transform (FFT), OTFS converts time-varying channels into invariant channels in a domain called delay-Doppler (also known as the Zak domain). 
Through the OTFS waveform, all symbols in a transmission block experience the same channel gain, providing a high order of diversity. In this way, OTFS is more robust to the carrier frequency shift, being more suitable for use in high mobility scenarios and millimeter wave systems.  Also, depending on the block size to be transmitted, the PAPR of OTFS is much smaller than the one from OFDM and FBMC.
The OTFS technique uses the intrinsic complex orthogonality of the OFDM transceiver to convert a time-dispersive channel to a delay-Doppler invariant channel. Thus, in order to change the OFDM transceiver to another one based on a filter bank it is necessary to guarantee the complex orthogonality of the last one. In \cite{zakaria2012novel} and \cite{demmer2017block} time spreading via FFT was proposed in order to recover the complex orthogonality. Through a system called FFT-FBMC, the authors aim to eliminate the intrinsic filter interference. In \cite{nissel2018pruned}, the authors present a precoding technique based in a pruned Discrete Fourier Transform (DFT) combined with a filter compensation scheme.
This scheme, called pruned DFT spread FBMC, presents the advantages of FBMC modulation and single-carrier frequency division multiple access (SC-FDMA), such as good spectral localization and low PAPR, in addition to partial restoration of the complex orthogonality.
However, to eliminate the intrinsic filter interference it is necessary to use a prototype filter with an overlapping factor lower or equal to 1.5.
This generates higher OOB emissions when compared to conventional FBMC/OQAM and FFT-FBMC systems, which do not have this restriction. 
In \cite{pereira2020novel} a system that shares some characteristics, such as partial complex orthogonality, with Pruned DFT Spread FBMC was proposed. This scheme, called DFT precoded filter bank, also uses the double rate transmission principle but with a data transmission strategy without the use of OQAM. In addition, a bespoke iterative block decision feedback equalizer (IB-DFE) is also introduced in order to obtain better performance in terms of bit error rate (BER) and in some scenarios, allow higher filter overlap factors and consequently a better spectral localization. Finally, a generalization of the DFT precoded filter bank was proposed in \cite{pereira2022generalized} by changing the waveform structure. This  obtains an even lower PAPR and robustness to high mobility scenarios.

In this work, a new OTFS system based on the DFT precoded filter bank system transceiver is proposed. 
The idea is to combine a system with good spectral localization, complex orthogonality and high robustness in high mobility scenarios. A common matrix structure is presented to understand the proposed waveform and compare it with other systems. 
It is shown that the interference can be greatly reduced in high spread Doppler environments. The main contributions of this work are:
\begin{enumerate}
    \item A new precoded filter bank scheme with complex orthogonality, robustness to double selective channels and lower PAPR and OOB emissions;
    \item A hybrid receiver with good performance and lower complexity than many equalizers in the delay-Doppler domain;
        \item A mathematical analysis of the new scheme via a simple matrix representation;
    \item A performance evaluation comparing the proposed system with other multi-carrier systems.
       \end{enumerate}

The remainder of this paper is organized as follows:  Section II describes the filter bank system as well as the process of restoring the complex orthogonality via precoding of this system. Section III presents the proposed system and highlights its advantages in some application scenarios. Section IV analyzes the system in doubly selective channels. 
Numerical results are presented in Section V; finally, Section VI concludes the article and presents some future perspectives. 

\textit{Notation}: vectors and matrices are represented by lowercase and uppercase letters in bold, respectively. The superscripts $(.)^T$ and $(.)^H$ denote, respectively, transpose, and Hermitian transpose operations. The identity matrix of size $N \times N$ is denoted by $\mathbf{I} _N$.
We will use $[\mathbf{M}]_{i,j}$ to refer to the (i;j)-th element of a matrix $\mathbf{M}$, and $[\mathbf{m}]_{i}$ to i-th element of the vector $\mathbf{m}$. The diag($\mathbf{M}$) produces a new vector with the same elements as the main diagonal of $\mathbf{M} $ and diag($\mathbf{m}$) represents the generation of a diagonal matrix of the elements of the vector $\mathbf{m}$. The column vectorization of a matrix is represented by vec(.)

 \section{Orthogonal Time-Frequency Space description}

The OTFS modulation has been introduced by Hadani et al.  \cite{hadani2017orthogonal} as a promising technique to combat the Doppler effect produced by doubly dispersive time-varying  multipath  channels. 

At the transmitter, the symbols of a QAM constellation positioned on a delay-Doppler grid are mapped to symbols of a time-frequency grid by the inverse SFFT (ISFFT) and transmitted over the channel using the OFDM multi-carrier technique \cite{murali2018otfs}. Basically, it can be seen as a precoding via SFFT in a system that uses an OFDM transceiver as its core. 
Such a technique converts the time-dispersive channel into an invariant channel within the delay-Doppler domain. Thus, combined with a sophisticated equalization scheme in this delay-Doppler domain the information symbols inside the block undergo almost constant attenuation \cite{hadani2018otfs}.

\begin{figure}[!t]
\centering
\includegraphics[width=0.48\textwidth]{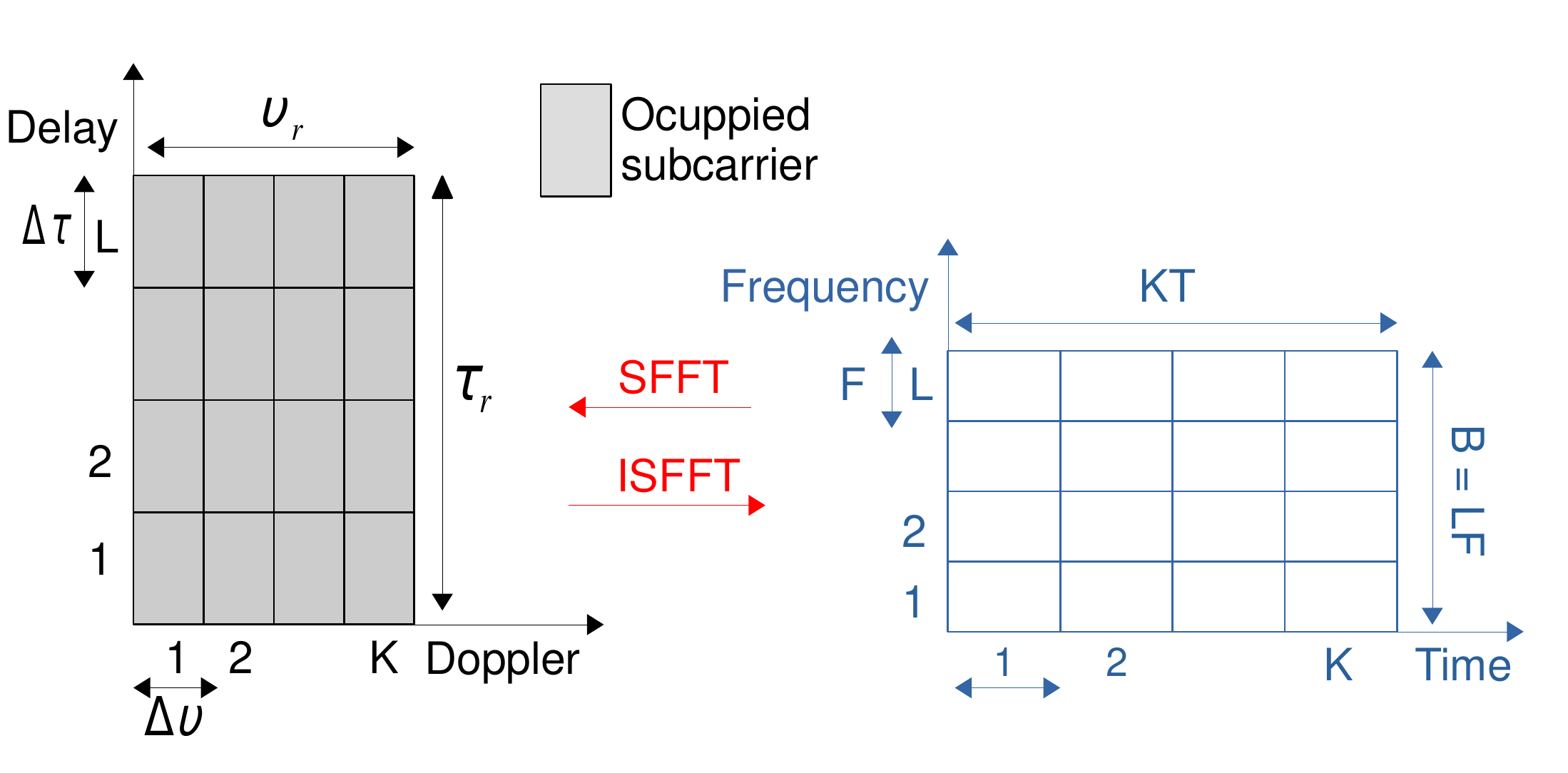}
\caption{Symplectic Fourier Duality - Analysis of time-frequency and delay-Doppler grids.}
\label{grade}
\end{figure}

A multi-carrier system can be characterized as a block structure consisting of $K$ symbols with a duration of $T$ seconds, where each symbol has $L$ subcarriers spaced from each other by $F$ Hz. 
Thus, the time-frequency grid consists of $L$ points spaced by $F$ along the frequency axis and $K$ points along the time axis with $T$ spacing, as shown in Figure \ref{grade}.
In this sense, we can define the bandwidth $B = LF$ and the total transmission time interval as $KT$. The delay axis length is given by the delay spread $\tau_r = 1/F$, while the Doppler axis length is given by the maximum Doppler shift $\upsilon_v = 1/T$. Thus, the Doppler-delay grid consists of $L$ points along the delay domain with $\Delta \tau = \frac{1}{LF}$ spacing and $K$ points along the Doppler domain with $\Delta \nu = \frac{1}{KT}$ spacing. Consequently, the transmission bandwidth $B$ is the inverse of the delay resolution and the total transmission duration $KT$ is the inverse of the Doppler resolution. 

Figure \ref{grade} presents the respective grids from the perspective of the SFFT and the ISFFT. Basically, while in a conventional OFDM or FBMC system the information data is represented from a time-frequency grid, in OTFS the data initially comes from a delay-Doppler grid. Thus, through the ISSFT we retrieve these data in the time-frequency grid for its eventual transmission by the specific multi-carrier technique. 
The SFFT is a two-dimensional Fourier transform; in this sense, it is equivalent to implement a $K$-point IFFT followed by an $L$-point FFT. Based on these definitions and using a matrix notation to formulate the system model, the signal at the input of the OFDM modulator $\mathbf{X}_{\mathrm{OTFS}} \in \mathbb{C}^{L \times K}$ is given by 
\begin{eqnarray}
\mathbf{X}_{\mathrm{OTFS}} = \mathbf{W}_L \mathbf{A} \mathbf{W}^H_K,
\label{eq:x_otfs}
\end{eqnarray}
where $\mathbf{A} \in \mathbb{C}^{L \times K}$ represents the transmitted complex symbols in the delay-Doppler domain and $\mathbf{W}_n = \big\{e^{j2\pi kl/n}  \big\}^{n-1}_{k,l = 0}   \in \mathbb{C}^{n \times n}$ is the $n$-point DFT matrix. 
After the precoding process, we implement the conventional OFDM modulation through an $L$-point IDFT. Thus, disregarding the cyclic prefix the transmitted signal can be expressed by:
\begin{eqnarray}
\mathbf{S}_{\mathrm{OTFS}} = \mathbf{W}^H_L \mathbf{X}_{\mathrm{OTFS}} =  \mathbf{A} \mathbf{W}^H_K
\label{OFDM_OTFS}
\end{eqnarray}
The column vectorization of $\mathbf{S}_{\mathrm{OTFS}} \in \mathbb{C}^{L \times K}$ in~\eqref{OFDM_OTFS} yields $\mathbf{s}  \in \mathbb{C}^{LK \times 1} = \textrm{vec}(\mathbf{S}_{\mathrm{OTFS}}) = (\mathbf{W}^H_K  \otimes \mathbf{I}_L ) \mathbf{a}$, where $\mathbf{a} \in \mathbb{C}^{LK \times 1} = \textrm{vec}(\mathbf{A})$ and  $\otimes$ refers to the Kronecker product.
At the receiver the inverse process is applied. First, OFDM demodulation is performed through the DFT returning the signal to the frequency domain. 
Finally, through a SFFT we resume the symbols transmitted in the delay-Doppler domain for detection. 
Figure \ref{otfs235t} illustrates the complete OTFS system.
{
\selectlanguage{english}
\begin{figure}[!b]
 \centering
\includegraphics[width=0.48\textwidth]{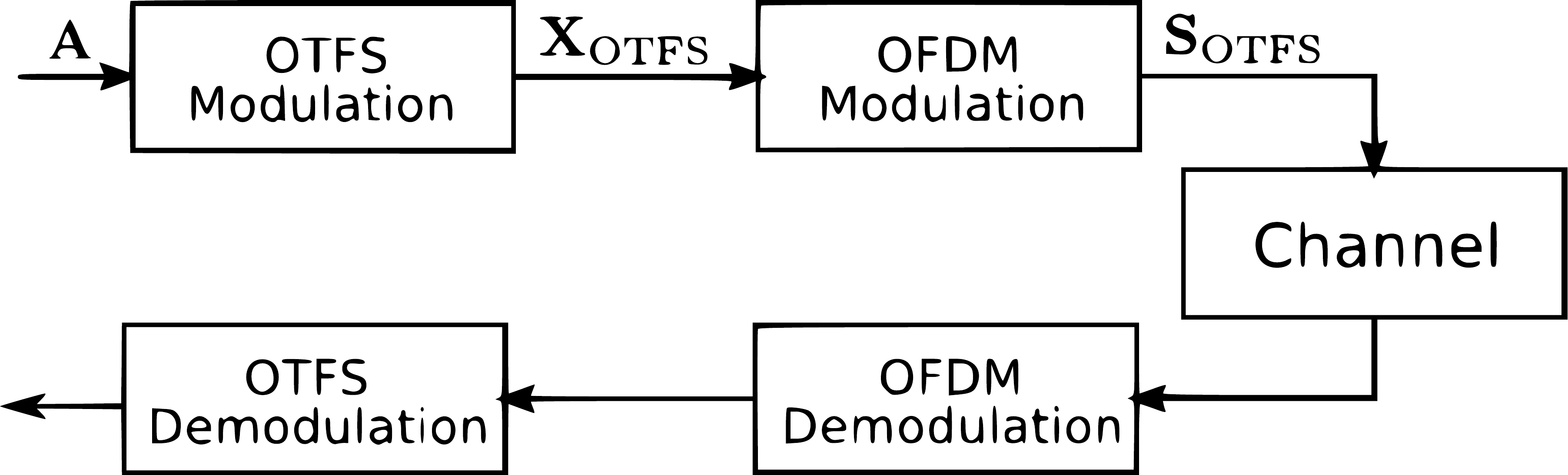}
\caption{Block diagram of an OTFS System.}
\label{otfs235t}
\end{figure}
}

Due to the use of a OFDM modulator, the OTFS technique presents significative OOB emissions, which are inherent to OFDM. Filter bank-based systems, such as FBMC/OQAM, aim to minimize this problem through a filtering process on each subcarrier. However, this filtering generates interference and consequently the loss of complex orthogonality. In the next section, we will present a new waveform with complex orthogonality based on the OTFS combined with a filter bank, which we call 2D-FFT FB.

\section{2D-FFT FB Scheme}

\begin{figure*}[h!]
 \centering
\includegraphics[width=0.99\textwidth]{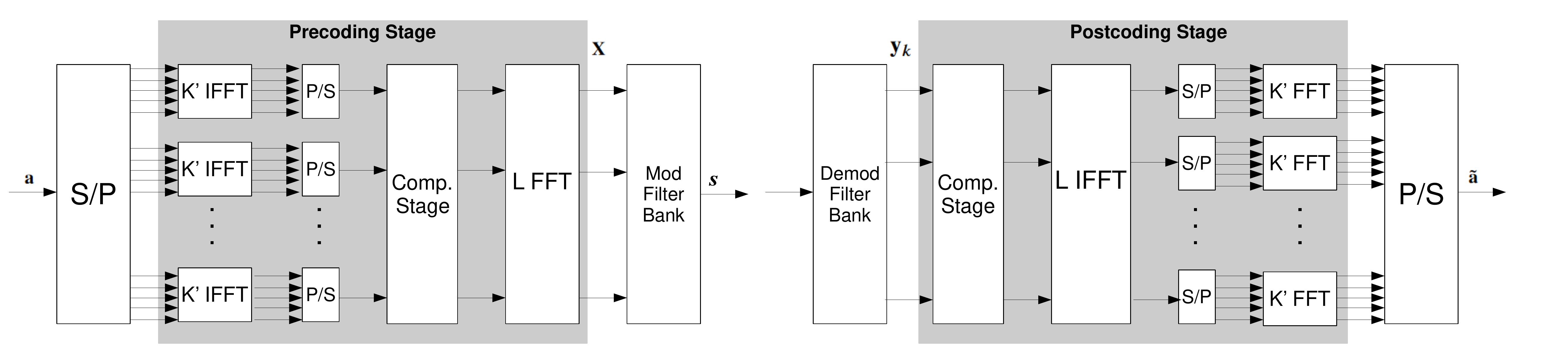}
\caption{Block scheme of the 2D-FFT filter bank system.}
\label{systemds}
\end{figure*}

The OTFS technique applies the OFDM multi-carrier system as the core for transmission. However, other multi-carrier techniques can also be used as a kernel to implement this transmission, as long as complex orthogonality between the pulses of the transmitter/receiver is guaranteed \cite{hadani2018otfs}. Figure \ref{systemds} shows the OTFS technique using a filter bank multi-carrier scheme.
Basically, there is an IFFT performed in the Doppler domain and a FFT in the delay domain, generating a time-frequency spreading via ISFFT as in OTFS. The difference lies in the addition of a compensatory multiplicative factor and in the structure of the data. Furthermore, after mapping the symbols on the time-frequency grid these will now be transmitted by a filter bank. 

Let us consider $x_{l,k}$ as a complex transmitted symbol in the $l$-th subcarrier at a time instant $k$. The baseband equivalent of a discrete time filter bank signal can be expressed as \cite{siohan2002analysis}
\begin{eqnarray}
s[m] =  \sum_{l=1}^{L} \sum_{k=1}^{K'} x_{l,k} g_{l,k}[m],
\label{25d145}
\end{eqnarray}	
where $K' = 2K$, $g_{l,k}[m]$ is given by
\begin{eqnarray}
g_{l,k}[m] = g[m-kL/2] e^{j\frac{2\pi l}{L}(m-kL/2)}
\label{25d14d5}
\end{eqnarray}	
and is essentially a time and frequency-shifted version of the prototype filter $g[m]$. The pulse used in the filtering process has unit energy and has a length of $OL$, where $O$ denotes the filter's overlap factor. Note that the pulse duration is controlled by the overlap factor $O$.

Normally, while OFDM has complex orthogonality between pulses with maximum symbol density ($TF = 1$), FBMC/OQAM systems have time-frequency spacing $TF = 0.5$. This generates intrinsic imaginary interference, restricting the orthogonality to the real field. However, complex orthogonality can be restored through precoding techniques, by spreading the symbols in time or frequency \cite{zakaria2012novel}.
The authors in  \cite{nissel2018pruned} and \cite{pereira2020novel} have proposed a precoding technique based on frequency spreading via DFT combined with a filter compensation stage.
Basically, $L/2$ complex data symbols are spread over $L$ subcarriers, leading to the same information rate as conventional FBMC/OQAM. Thus, the data symbols no longer belong to a position on a specific frequency, but are spread over several subcarriers. 
Particularly in \cite{pereira2020novel},
$L/2$ complex symbols are properly precoded to guarantee the complex orthogonality without using OQAM symbols, as will be explained later. Thus, 
through this technique combined with the IFFT blocks, an ISFFT is implemented and therefore, based on~\eqref{eq:x_otfs} the signal at the filter bank input represented by the matrix $\mathbf{X} \in C^{L \times K'}$, can be expressed by:
\begin{eqnarray}
\mathbf{X} = \mathbf{W}_L \textrm{diag}\{\tilde{\mathbf{b}}\} \mathbf{A}'\mathbf{W}^H_{K'}
\label{OTFSFB}
\end{eqnarray}
where $\mathbf{A'} \in C^{L \times K'}$ contains $K'$ multi-carrier symbols, each with $L/2$ complex data symbols from a QAM constellation. These symbols are positioned in the first and last $L/4$  positions of symbols with zeros in the remaining positions. Such a transmission strategy is necessary to avoid interference in the frequency domain \cite{pereira2020novel}. The vector $\mathbf{\tilde{b}} \in C^{L \times 1}$ represents the filter compensation stage used in the complex orthogonality recovery process that will be derived later in this work.
The use of two times more multi-carrier symbols ($K'$) with respect to the conventional OTFS system is necessary so that a fair comparison with OTFS can be done in order to obtain the maximum data transmission capacity. However, we will see that this will not impact on a better discrimination of the delay-Doppler domain by our system since the transmission is done at a double rate through the filter bank.
There is a certain similarity between Equations (\ref{eq:x_otfs}) and (\ref{OTFSFB}), since they represent the samples that will be transmitted by the specific multicarrier modulator. While (\ref{eq:x_otfs}) uses the OFDM technique, in (\ref{OTFSFB}) a filter bank system will be used.
The addition of the compensation stage, as well as the structure of symbols from the QAM constellation, are also noteworthy.
In general, it is possible to see in both cases a coding based on a ISFFT on the data symbols; the difference is in the multicarrier core mentioned above.

The filter bank system can be efficiently implemented through a prototype filter decomposition via a polyphase network (PPN) combined with an IDFT \cite{siohan2002analysis}. Thus, initially applying an $N$-point IDFT, we produce the matrix $\mathbf{D} \in C^{N \times K'}$ which can be expressed,
\begin{eqnarray}
\mathbf{D} = \mathbf{\tilde{W}}^H_{N} \mathbf{X},
\label{sfsadf}
\end{eqnarray}
where $\mathbf{\tilde{W}}_{N} \in C^{L \times N}$ is a DFT matrix which is given as follows
\begin{eqnarray}
\mathbf{\tilde{W}}_{N} = \begin{bmatrix}
\mathbf{I}_L &  \mathbf{0}_{L\times N-L}  \end{bmatrix}
\mathbf{W}_N,  
\label{dft_espalhada}
\end{eqnarray}
that corresponds to spreading $L$ frequency data into $N$ time data, with $N > L$ so that the DFT of the precoding stage does not cancel itself out with the IDFT of the filter bank.
In this context, we will change the order of~\eqref{25d14d5} from $L$ to $N$ and the prototype filter has length $ON$. The transmitted data can be obtained by convolving $\mathbf{d}_k$ with the prototype filter impulse response through a Toeplitz filter matrix.
For this, let us consider the diagonal matrix $\textbf{G}_o$ corresponding to the filter coefficients,  that is, $\textbf{G}_o =$ diag$(\textbf{g}_o) \in R^{N/2 \times N/2}$ for $o = 0,1,2,...,2O-1$, where $\mathbf{g}_{o}$ is as follows 
$\mathbf{g}_o = [g[oN/2],g[oN/2+1],   \ldots , g[oN/2+N/2-1]]$. 
Therefore, the Toeplitz matrix of the filter $\mathbf{G}$ $\in \mathbb{R}^{ON+(K'-1)N/2 \times NK'}$ can be given as follows:
\begin{eqnarray}
\mathbf{G} = 
\begin{bmatrix}
\mathbf{G}_0  & \mathbf{0} & \mathbf{0} & \mathbf{0} & \ldots  & \mathbf{0} \\
\mathbf{0} & \mathbf{G}_1  & \mathbf{G}_0  & \mathbf{0} & \ldots  & \mathbf{0} \\
\mathbf{G}_2  & \mathbf{0} & \mathbf{0} & \mathbf{G}_1  &  \ldots  & \mathbf{0} \\
 \mathbf{0} & \mathbf{G}_3  & \mathbf{G}_2 & \mathbf{0} & \ldots  & \mathbf{0} \\
 \vdots &  \mathbf{0} &  \mathbf{0} & \mathbf{G}_3  &  \ldots  & \mathbf{0} \\
 \vdots &  \vdots &  \vdots &  \vdots &  \ddots  &   \vdots \\
 \mathbf{G}_{20-4} & \vdots &  \vdots &  \vdots & \ddots  &  \mathbf{0} \\
 \mathbf{0} & \mathbf{G}_{2O-3} & \mathbf{G}_{20-4} &  \vdots & \ddots & \mathbf{G}_1 \\
 \mathbf{G}_{2O-2} &  \mathbf{0} &  \mathbf{0} &  \mathbf{G}_{2O-3} & \ddots &  \mathbf{0} \\
 \mathbf{0} & \mathbf{G}_{2O-1} & \mathbf{G}_{2O-2} & \mathbf{0} & \ddots &  \mathbf{G}_3 \\
 \vdots &  \mathbf{0} & \mathbf{0} & \mathbf{G}_{2O-1} & \ddots & \vdots \\
  \vdots &  \vdots & \ddots &   \vdots &  \ddots &  \mathbf{0} \\
   \mathbf{0} &   \mathbf{0} &  \ldots & \mathbf{0} &  \ddots & \mathbf{G}_{2O-1} 
\end{bmatrix}.
\label{fhddg}
\end{eqnarray}
Thus, using the double rate strategy of FBMC/OQAM systems, the output data vector $\mathbf{s}$ of length $M = ON + (K'-1)\frac{N}{2}$ from the filter bank is given by \cite{pereira2020novel}:
\begin{eqnarray}
 \mathbf{s} = \mathbf{G}\mathbf{d}
 \label{s_FB}
\end{eqnarray}
where $\mathbf{d} = \textrm{vec}(\mathbf{D}) \in C^{NK' \times 1}$  are the symbols in seralized form. Thus, through the proposed matrix $\mathbf{G}$, we have $L/2$ complex symbols with twice the transmission rate equivalent to the $L$ complex symbols of an OFDM transmission.

The transmission of $\mathbf{s}$ is affected by additive white Gaussian noise (AWGN) and a time-varying multipath channel represented in the convolution matrix $\mathbf{H} \in C^{M \times M}$ \cite{nissel2018pruned}. Thus, the received signal $\mathbf{r} \in C^{M \times 1}$ can be described by
\begin{eqnarray}
 \mathbf{r} = \mathbf{H}\mathbf{s} + \mathbf{n},
\end{eqnarray}
where $\mathbf{n}$ are the AWGN samples with zero mean and power $\sigma_n^2$. Similarly, in the receiver the reverse process of the transmitter is done. Initially, the received signal is demodulated by the analysis filter $\mathbf{G}^T$. The demodulated signal $\mathbf{z} \hspace{0.1cm} \in C^{NK' \times 1}$ can be expressed as follows:
\begin{eqnarray}
 \mathbf{z} &=&  \mathbf{G}^T \mathbf{r}
\label{sdgfd246}
\end{eqnarray}
The detected vector $\mathbf{y}_k \in C^{L\times 1}$ is obtained after a $N$-point DFT as
\begin{eqnarray}
\mathbf{y}_k = \mathbf{\tilde{W}}_{N}  \mathbf{z}_k.
\label{sdgsd780}
\end{eqnarray}
To compensate for interferences coming from the channel, a  one-tap equalization process is performed by $\mathbf{e}_k \hspace{0.1cm} \in C^ {L \times 1}$, originating $\mathbf{\tilde{x}}_k\in C^{L\times 1}$ which is expressed as 
\begin{eqnarray}
\mathbf{\tilde{x}}_k=\textrm{diag}\{\mathbf{e}_k\}\mathbf{y}_k.  
\label{equali}
\end{eqnarray}
Finally and considering all time instants, the detected symbols represented by $\mathbf{\tilde{A}} \in C^{L \times K'}$ are obtained through the SFFT combined with the compensation stage:
\begin{eqnarray}
\mathbf{\tilde{A}} = \mathbf{W}^H_L \textrm{diag}\{\tilde{\mathbf{b}}\} \mathbf{\tilde{X}}\mathbf{W}_{K'}
\label{12ghtj}
\end{eqnarray}
where $\mathbf{\tilde{X}} = [\mathbf{\tilde{x}}_1;\mathbf{\tilde{x}}_2;\hdots;\mathbf{\tilde{x}}_K'] \in C^{L \times K'}$. To obtain the estimated symbols, the $L/2$ intermediate values between the first and last $L/4$ symbols are discarded, since no data was transmitted in those positions in $\mathbf{A}$. We can see that while OTFS looks like a generalization of SC-FDMA, the  2D-FFT is a generalization of the filter bank systems with spreading via DFT proposed in \cite{nissel2018pruned} and \cite{pereira2020novel}.
\begin{figure}[!t]
\centering
\includegraphics[width=0.48\textwidth]{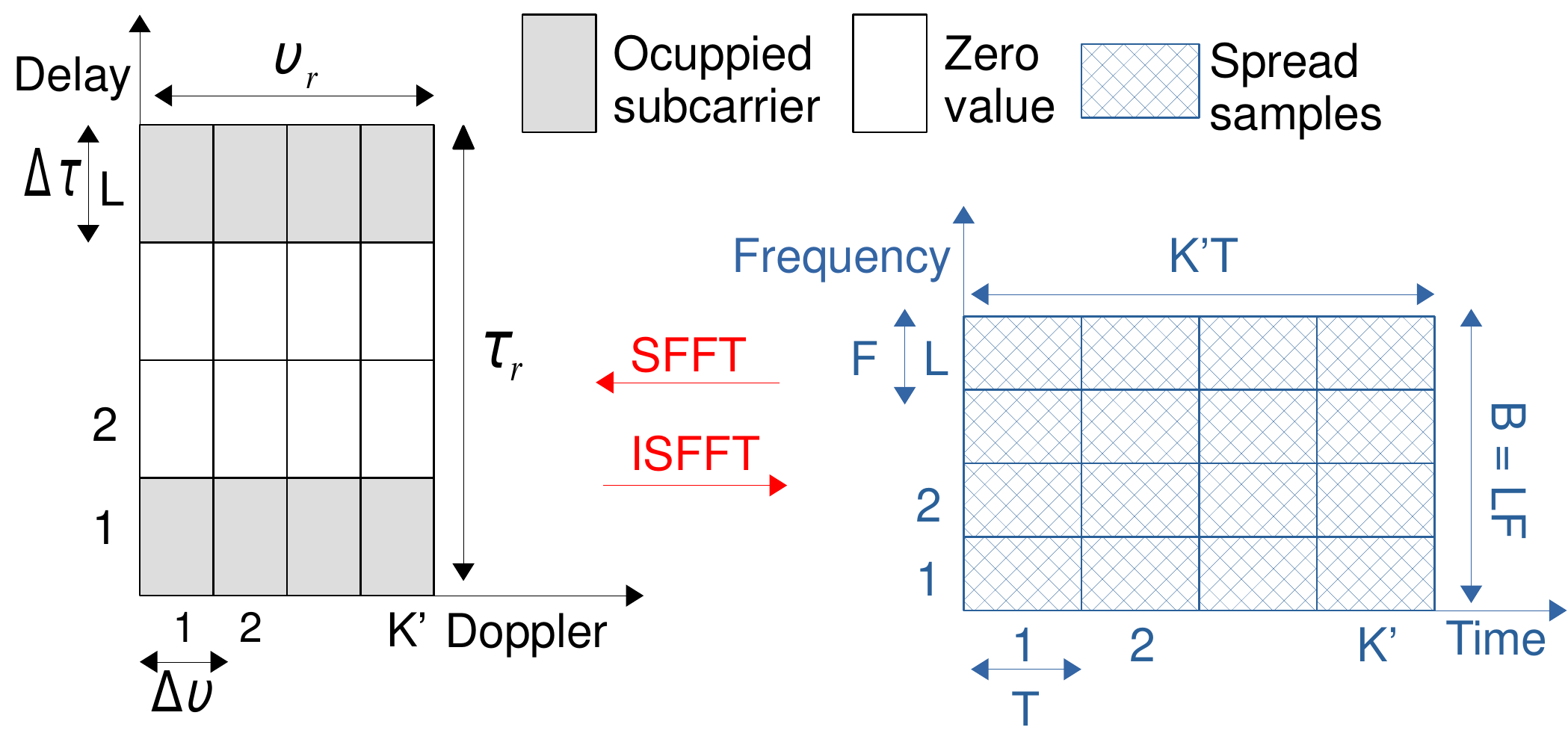}
\caption{Analysis of time-frequency and delay-Doppler grids in the proposed scheme.}
\label{grade3}
\end{figure}

To use OTFS with a FBMC transceiver as its core it is necessary to eliminate the interference coming from the filters in order to recover complex orthogonality. We will see below that this can be done using a part of the OTFS itself with the addition of a multiplicative factor of compensation. 

\subsection{Complex orthogonality restoration}

To establish complex orthogonality in a FBMC system we will use a precoding method based on DFT spreading combined with a pre-compensation stage, as proposed in \cite{pereira2020novel}. This transmission technique does not use OQAM modulation. The principle of this scheme can be summarized as follows: $L/2$ complex symbols are pre-compensated by a scale factor and are inserted in the first and last $L/4$ positions of a $L$-point DFT, with the intermediate values equal to zero in order to spread the symbols over all subcarriers.

The compensation stage is a multiplicative factor based on the prototype filter coefficients. Thus, it is necessary that the interference from the filter overlapping is only one coefficient. As explained in \cite{nissel2018pruned,pereira2020novel}, this goal is achieved by constraining the overlapping factor $O$ to a value less than or equal to $1.5$. Assuming such a limitation, the coding/spreading matrix defined here by $\mathbf{C}_f$ must establish the following condition \cite{pereira2020novel}:
\begin{eqnarray}
 \mathbf{C}^H_f\mathbf{\tilde{W}}_{N}\mathbf{\widetilde{G}}^T\mathbf{\widetilde{G}}\mathbf{\tilde{W}}^H_{N} \mathbf{C}_f \approx 
 \mathbf{F}, 
 \label{sdvsd}
\end{eqnarray}
where $\mathbf{\widetilde{G}} \in R^ {ON \times N}$ refers to  $\mathbf{G}$ for $K' = 1$, that is, it corresponds to the transmission of a single multi-carrier symbol. For this, the first two columns of the matrix are used with a number of rows equal to $ON$.The matrix $\mathbf{F} \in  R^ {L \times L}$ is the expected result of recovering the complex orthogonality with ones on the first and last $L/4$ positions of its main diagonal and zeros elsewhere.
As commented, a coding that fulfills this condition is one that applies a DFT on the transmitted symbols combined with a filter compensation stage, that is,
\begin{eqnarray}
 \mathbf{C}_f = \mathbf{W}_{L} \textrm{diag}\{\tilde{\mathbf{b}}\},
 \label{ferf4}
\end{eqnarray}
Note that, if~\eqref{eq:x_otfs} and ~\eqref{OTFSFB} are analyzed, $\mathbf{C}_f$ is already implemented, because $\mathbf{W}_{L}$ is part ISFFT of the 
used to implement the OTFS technique.
Therefore, it is only necessary to add the vector $\tilde{\mathbf{b}}$ that has $L/2$  compensation coefficients in the positions according to the transmission strategy defined for $\mathbf{A}$, with its $L/2$ intermediate samples equal to zero.
Replacing~\eqref{ferf4} in~\eqref{sdvsd}, the $i$-th position of $\tilde{\mathbf{b}}$ can be expressed by
\begin{eqnarray}
[ \mathbf{\tilde{b}}]_i = \sqrt{\frac{1}{[\mathbf{\tilde{c}}]_i}},\hspace{0.2cm} \textrm{for} \hspace{0.2cm} i = \left[ 1,\ldots,\frac{L}{4} ;   L-\frac{L}{4},\ldots,L \right],
\end{eqnarray}
with
\begin{eqnarray}
 \mathbf{\tilde{c}} &=& \textrm{diag}\{\mathbf{W}^H_L\mathbf{\widetilde{W}_N}\mathbf{\widetilde{G}}^T\mathbf{\widetilde{G}}\mathbf{\widetilde{W}_N}^H\mathbf{W}_L\}. 
\end{eqnarray}
Due to the product $\mathbf{\widetilde{G}}^T\mathbf{\widetilde{G}}$, it can be seen that this compensation coefficient depends only on the prototype filter response (base pulse) which is constant over time. Thus, this coefficient can be applied by a point-to-point multiplication on all active subcarriers. Therefore, $\mathbf{C}_f$ is multiplied with the data vector of size $L$ containing $L/2$ complex symbols in their positions according to the pre-established transmission strategy. 
Although the time-frequency spacing (density) is equal to
$TF=0.5$, only $L/2$ complex symbols are transmitted, leading to an equivalent time-frequency spacing of $TF=1$ for $L$ complex symbols.

Frequency-Time spreading is interesting in filter bank systems because it is the basis for solving the problem of complex orthogonality loss.
Through spreading, we have an additional code dimension (in addition to time and frequency) that allows us to eliminate the interference imposed by the filter at the time of transmission and reception. Therefore, such coding is done both at the transmitter and at the receiver. By recovering complex orthogonality, equalization becomes simpler as the symbols are now free from filter interference. Furthermore, it is possible to see from the signal~\eqref{OTFSFB} at the input of the filter bank that OTFS can naturally be used, since the orthogonality recovery process is a part of the OTFS. The novelty in our proposal is that $L/2$ information symbols are transmitted at each time instant at a doubled rate, as well as using a filter compensation stage.
It is interesting to note that in a conventional OTFS transmission a delay-Doppler grid ($L \times K$) is transformed into a time-frequency grid ($K \times L$) through the inverse symplectic transform (Figure \ref{grade}). In our proposed technique, a grid ($L \times K'$) in the delay-Doppler domain, with only $L/2$ active subcarriers, is transformed into a time-frequency domain grid ($K' \times L$), as shown in Figure \ref{grade3}. To maintain the same symbol density of a ($K \times L$) conventional OTFS (for comparison purposes), we must have $K'=2K$. So, the grid ($2K  \times L$) is divided in two grids ($K \times L$) and transmitted at a doubled rate. This transmission technique leads our technique to have the same Doppler spread discrimination as a ($K \times L$) conventional OTFS.

\subsection{Computational complexity}

Table \ref{tabelacodm} presents a comparison of transmission scheme with other techniques mentioned in this work in terms of computational complexity of the transmitter. If we ignore the channel equalization process, the table also represents the complexity of the receiver.
\begin{table}[!ht]
\centering
\scalebox{0.9}{
\begin{tabular}{|c|c|}
\hline
Scheme            & Complexity                      \\ \hline
FBMC              & $L + N \log N + ON$           \\ \hline
DFT Precoded FB &  2 ($\frac{L}{2} + L\log (L/2) + N\log N + ON$) \\ \hline
OTFS &  $LK\log (L) + LK\log (K)   + L\log L      $\\ \hline
2D FFT-FB & $\frac{L}{2} + LK'\log (L/2) + (L/2)K'\log (K')   +     N\log N + ON      $ \\ \hline
\end{tabular}}
\caption{Computational complexity of systems.}
\label{tabelacodm}
\end{table}
We recall that the DFT can be implemented efficiently by a FFT, where the number of complex multiplications for an $L$-point FFT is $L$log$_2 L$. Thus, the term $N\log N + ON$ corresponds to the inverse FFT and the $ON$ multiplications refer to the prototype filter needed in all schemes. Frequency spreading via FFT, present in DFT Precoded FB and 2D FFT-FB, has a complexity of $L\log L$ multiplications. Both schemes also have $L/2$ multiplications referring to the scaling factor for filter pre-compensation. On the other hand, time spreading (also via IFFT) has a
complexity of $K'\log K'$ multiplications and is present in the system proposed in this work. Finally, for the case of FBMC, $L$ multiplications are necessary due to the phase factor that generates the $\pi/2$ rotation  \cite{siohan2002analysis}.

The OFDM transceiver is the simplest way to implement the OTFS transmission, since  the condition of complex orthogonality between the transmitter/receiver pulses is implicitly guaranteed \cite{hadani2017orthogonal}. For regular FBMC systems, the loss of complex orthogonality makes the use of OTFS unfeasible. However, as previously seen complex orthogonality is guaranteed in the proposed scheme through precoding techniques defined by $\mathbf{C}_f$ and by limiting the overlap factor of the prototype filter to $O \leq 1.5$.
The implementation of SFFT is easily performed by adding IDFT/DFT blocks in the precoding process ($\mathbf{C}_f$). We call this new transmission technique that generalizes SC-FDMA and FBMC systems as 2D-FFT-FB. Such a proposal has the following advantages over other multi-carrier techniques: 
\begin{enumerate}
\item Does not use a cyclic prefix; 
\item Lower OOB due to filtering process; 
 \item PAPR similar to OTFS;
 \item Robustness to double selective channels;
 \item Allows the usage of simple frequency domain equalizers with good performance.
\end{enumerate} 
The main objective of the proposed system is to improve the performance in high mobility scenarios and to minimize OOB emissions. Moreover, as it will be shown in the following sections 2D-FFT-FB has good performance with low complexity frequency domain equalizers.

\section{Receivers for 2D-FFT FB}
\label{section5}

The effect of channel interference can be reduced by proper signal processing method on the receiver side. Generally, in OFDM and FBMC, one-tap equalizers in frequency domain calculated using the minimum mean squared error (MMSE) or zero forcing (ZF) criteria are used to compensate these interferences. In the OTFS structures, receivers in the delay-Doppler domain usually have a higher degree of complexity. In this section, we present a one-tap equalizer in the frequency domain and its MMSE version in the delay-Doppler domain. Finally, a two-stage iterative hybrid receiver is proposed to minimize the effects of the dual selective channel. In the first stage, the conventional one-tap MMSE equalizer in the frequency domain is used, bringing reliable estimates that will be fed to an iterative interference canceller in the delay-Doppler domain, obtaining a good final estimate of the symbols. It should be noted that the use of a doubled rate transmission associated with the precoding technique of the proposed system positively impacts the system performance with respect to the Doppler effect. More specifically, the one-tap equalizer in the frequency domain achieves a sufficiently good performance that when combined with interference cancellation in the delay-Doppler domain it provides a better BER performance when compared to other techniques that use conventional OTFS modulation, as it will be demonstrated by simulation results.

\subsection{Frequency Domain One-Tap Equalizer}
\label{subsec_freq_eq}

In order to obtain the optimal channel coefficients in frequency domain, we will consider the global matrix notation of the implemented filter bank system. For this, we are not interested in just a specific time $k$, but in the entire transmitted block. Thus, we define the DFT matrix $ \mathbf{\tilde{W}} \in C^{LK'\times NK'}$ as
\begin{eqnarray}
 \mathbf{\tilde{W}} =  \mathbf{I}_{LK'} \otimes \mathbf{\tilde{W}}_N,
\end{eqnarray}
which basically maps $\mathbf{\tilde{W}}_N$ to the correct time positions. Hence, it is possible to rewrite~\eqref{s_FB} as  
\begin{eqnarray}
 \mathbf{s} = \mathbf{G} \mathbf{\tilde{W}}^H\mathbf{x} = \mathbf{\bar{G}}\mathbf{x} \label{transmission_global}
\end{eqnarray}
where $\mathbf{\bar{G}} = \mathbf{G} \mathbf{\tilde{W}}^H \in \mathbb{C}^{M \times LK'}$ and $\mathbf{x}  = \textrm{vec}(\mathbf{X}) \in \mathbb{C}^{LK' \times 1}$. One can represent the channel coefficients in the frequency domain from matrix product $\mathbf{\bar{G}}^H\mathbf{H}\mathbf{\bar{G}}$.
In many practical scenarios, the off-diagonal elements of $\mathbf{\bar{G}}^H\mathbf{H}\mathbf{\bar{G}}$ are so small that they are dominated by noise \cite{nissel2018pruned}. In this way, it is possible to represent the one-tap channel $\mathbf{h} \in C^{LK' \times 1 }$ by 
\begin{eqnarray}
 \mathbf{h} = \textrm{diag}\big\{\mathbf{\bar{G}}^H\mathbf{H}\mathbf{\bar{G}}\big\}.
\end{eqnarray}
Considering each position in time and frequency, we have
$h_{l,k}$ by
\begin{eqnarray}
h_{l,k} = [\mathbf{h}]_{l + L(k-1)}.
\end{eqnarray}
From this definition, a simple one-tap ZF equalizer is provided by $(1/h_{l,k})$. To obtain a balanced solution between noise and channel interference, we employ a MMSE equalizer, which can be expressed as follows: 
\begin{eqnarray}
e_{l,k} =  \frac{h^*_{l,k}}{ |h_{l,k}|^2 + \sigma_n^2}.
\end{eqnarray}
If delay propagation and Doppler effect are low enough, an one-tap equalizer is sufficient to compensate channel interference. Alternatively, more robust receivers can be used, which will improve performance especially in high mobility scenarios. In \cite{pereira2020novel} and \cite{junior2021iterative} iterative schemes were adopted that can strongly compensate for channel interference at the expense of greater receiver complexity.

\subsection{Delay-Doppler Domain Equalizer}

Frequency domain equalization is very useful in multi-carrier systems, such as OFDM and FBMC, because of its simplicity.
However, an OTFS system combined with an equalizer in the delay-Doppler domain is able to explore the full channel diversity  both in time and frequency.
Let us consider $\mathbf{H}_{ef} \in \mathbb{C}^{LK' \times LK'}$ as the effective channel matrix, that encompasses the entire transmission (filter bank) and coding process. So, unlike the frequency domain equalizer $\mathbf{e}_k$ which only considers the multi-carrier modulation stage, from now on  both the pre- and post-coding processes are also considered. Thus, $\mathbf{H}_{ef}$ can be written as:
\begin{eqnarray}
 \mathbf{H}_{ef} = \mathbf{C}^H \mathbf{\bar{G}}^H\mathbf{H}\mathbf{\bar{G}}\mathbf{C}, 
\label{input_output1}
\end{eqnarray}
where $\mathbf{C} \in C^{LK' \times LK'}$ corresponds to the system coding process for all transmitted blocks and is given by 
\begin{eqnarray}
 \mathbf{C} = (\mathbf{I}_{K'}  \otimes  \mathbf{C}_f) (\mathbf{W}_K'  \otimes  \mathbf{I}_{L}).
\end{eqnarray}
The MMSE equalizer in the delay-Doppler domain $\mathbf{E}_{dd} \in \mathbb{C}^{LK' \times LK'}$ can be expressed by 
\begin{eqnarray}
 \mathbf{E}_{dd} = \mathbf{H}^H_{ef}(\mathbf{H}_{ef}\mathbf{H}^H_{ef}       + \sigma^2_n\mathbf{I}_{LK'})^{-1}.
\label{fdfgdf}
\end{eqnarray}
Thus, the estimated symbols are given by 
\begin{eqnarray}
 \mathbf{\hat{a}} =  \mathbf{E}_{dd} \mathbf{\tilde{a}},
\end{eqnarray}
where $\mathbf{\tilde{a}} \in \mathbb{C}^{LK' \times 1} = \textrm{vec}(\mathbf{\tilde{A}})$. 
As we will see later in Section \ref{subsec_freq_eq}, OTFS technique combined with this delay-Doppler domain equalizer proves to be more robust in high mobility scenarios than when using a frequency domain equalizer. However, its complexity is significantly higher, mainly due to the need to invert an matrix of size $LK' \times LK'$. In the next subsection, a new hybrid receiver using a one-tap frequency domain equalizer associates to an iterative interference canceller in the delay-Doppler domain will be proposed.

\subsection{Hybrid Receiver For Interference Cancellation}

\begin{figure}[!t]
 \centering
\includegraphics[width=0.48\textwidth]{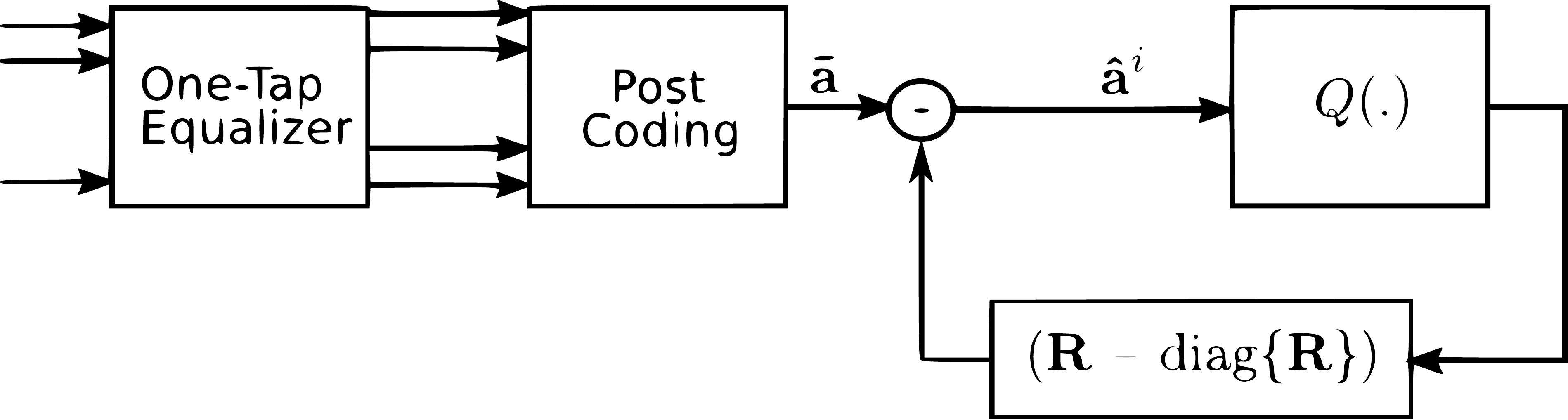}
\caption{Block scheme of the hybrid receiver.}
\label{IIC_Receiver}
\end{figure}

In order to handle more efficiently interferences stemming from the doubly-selective channel, Figure \ref{IIC_Receiver} presents a simple but effective interference cancellation scheme in an iterative format. For a very aggressive channel the first estimate may not be very accurate; so, detection schemes with interference estimation and cancellation are not always effective due to error propagation. Therefore, the challenge is to mitigate the error propagation through iterations, i.e., to improve the reliability of the symbols detected at each iteration. The proposed scheme contains a frequency domain MMSE equalizer followed by post-coding as its first stage, providing tentative estimates of the data vectors. Then, based on these estimates, the interference canceller calculates an estimate of the interference that should be minimized. 
Performing a parallel-serial conversion to $\mathbf{\tilde{A}}$ in~\eqref{12ghtj}, the input-output ratio of the entire transmission system $\mathbf{\tilde{a}}$ can be modeled by 
\begin{eqnarray}
\begin{matrix}  \mathbf{\tilde{a}} = \underbrace{\mathbf{C}^H \textrm{diag}\{\mathbf{e}\}\mathbf{\bar{G}}^H\mathbf{H}\mathbf{\bar{G}}\mathbf{C}}_{\mathbf{R}}\mathbf{a} + \underbrace{  \mathbf{C}^H \textrm{diag}\{\mathbf{e}\}\mathbf{\bar{G}}^H}_{\mathbf{\Upsilon}}\mathbf{n}.  \end{matrix} 
\label{input_output2}
\end{eqnarray}
The off-diagonal values of $\mathbf{R} \in C^{LK' \times LK'}$ and $\mathbf{\Upsilon} \in C^{LK' \times M}$ represent the interference induced by the channel and the noise, respectively. To reduce the level of interference and improve performance, the unwanted terms in $\mathbf{R}$ should be removed. 
Thus, similarly to what is suggested in \cite{junior2021iterative} it is possible to cancel the interference in the quantizer input for the $i$-th iteration described by $\mathbf{\hat{a}}^{i} \in C^{LK' \times LK'}$, according to
\begin{eqnarray}
 \mathbf{\hat{a}}^{i+1} = \mathbf{\tilde{a}} - (\mathbf{R} - \textrm{diag}\{\mathbf{R}\}) Q(\mathbf{\hat{a}}^i).
 \label{iccc}
\end{eqnarray}
where $Q(.)$ denoting the quantization operator. The algorithm is initialized with $\mathbf{\hat{a}}^0$ = $\mathbf{\tilde{a}}$, representing the system output using only the conventional one-tap equalizer $\mathbf{e}$.

In~\eqref{iccc}, $(\mathbf{R} - \textrm{diag}\{\mathbf{R}\})$
are the interference terms that are subtracted through
the iterative process, bringing greater reliability to the detected symbols.
Finally, looking closely at the receiver structure we see that there is no need to invert $\mathbf{R}$, unlike the MMSE delay-Doppler domain equalizer, which needs the inversion of  $\mathbf{H}_{ef}$. Thus, the proposed structure has a lower computational complexity, as seen in Table \ref{tabela2}. We recall that $\mathbf{H}_{ef}$ differs from $\mathbf{R}$ only by the exclusion of the frequency domain equalizer diag$\{\mathbf{e}\}$. 

In terms of complexity, Table~\ref{tabela2} presents the computational cost for the receivers presented in this section. 
\begin{table}[!t]
\centering
\begin{tabular}{|c|c|}
\hline
Receiver         & Complexity                       \\ \hline
MMSE - Frequency domain              & $LK'$           \\ \hline
MMSE - Delay-Doppler domain & $(LK'/2)^{3} + (LK'/2)^{2}$ \\ \hline
Hybrid receiver &  $LK' + (LK'/2)^2$\\ \hline
\end{tabular}
\caption{Computational complexity of the receivers.}
\label{tabela2}
\end{table}
The MMSE equalizer in the frequency domain, as it is a multiplicative factor of only one coefficient, has low complexity, requiring only $LK'$ multiplications.
For the MMSE equalizer in the delay-Doppler domain, we have the multiplication of an $LK' \times LK'$ matrix ($\mathbf{E}_{dd}$) by a vector ($\mathbf{\tilde{a}}$) of length $LK'$. The cost for this operation is $(LK')^2$ if we consider only multiplications. Furthermore, considering the Gauss Jordan elimination method for matrix inversion, more $(LK')^3$ multiplications are needed. However, we only have $LK'/2$ information data and the rest are zeros so we can then consider such a value for complexity. Finally, for the proposed hybrid receiver, $LK'$ multiplications of the one-tap equalizer in the frequency domain are necessary plus a cost of $(LK')^2$ by multiplying the matrix $\mathbf{R} -$ diag$\{\mathbf{R}\}$ by the vector $\mathbf{\hat{a}}$ if we consider only one iteration. Again, we can consider $(LK'/2)$ multiplications since $\mathbf{\hat{a}}$ has $(LK'/2)$ data and the remainder are zeros. It is worth mentioning that if the receivers are being applied in the OTFS technique, the value of $K'$ is substituted for $K$ and we will not have a division by 2 in the squared and cubed terms. 
The MMSE equalizer in the delay-Doppler domain has high computational complexity and should  be avoided in OTFS systems. There are other less computationally complex receivers proposed for OTFS such as those seen in \cite{zemen2017low,raviteja2018interference,tiwari2019low}. However, the study, comparison and implementation of these receivers in the proposed system are outside the scope of the work, and this is an interesting direction for future work.

\section{Numerical Results}
\label{sec:simresults}

To show the benefits of the proposed system in order to validate the work, this section present several simulation results. The simulation method chosen is Monte Carlo because it has a simple and very flexible structure. We start by presenting the PAPR and the spectral localization of the system compared to other techniques. Finally, we evaluated the error performance in terms of bit error rate (BER) in scenarios with medium delay spread and high mobility. The filter bank modulator (see Figure 3) used for the simulations was implemented using the Hermite filter detailed in \cite{haas1997time} which is based on a Gaussian function and therefore has a good location in the time-frequency plane. The Hermite pulse overlap factor is $O = 2$. Thus, we truncated the pulse to obtain $O = 1.5$ to avoid filter bank interference. The block duration of both systems is approximately the same.

\subsection{PAPR and spectral confinement}

\begin{figure}[!t]
\centering
\includegraphics[width=0.45\textwidth]{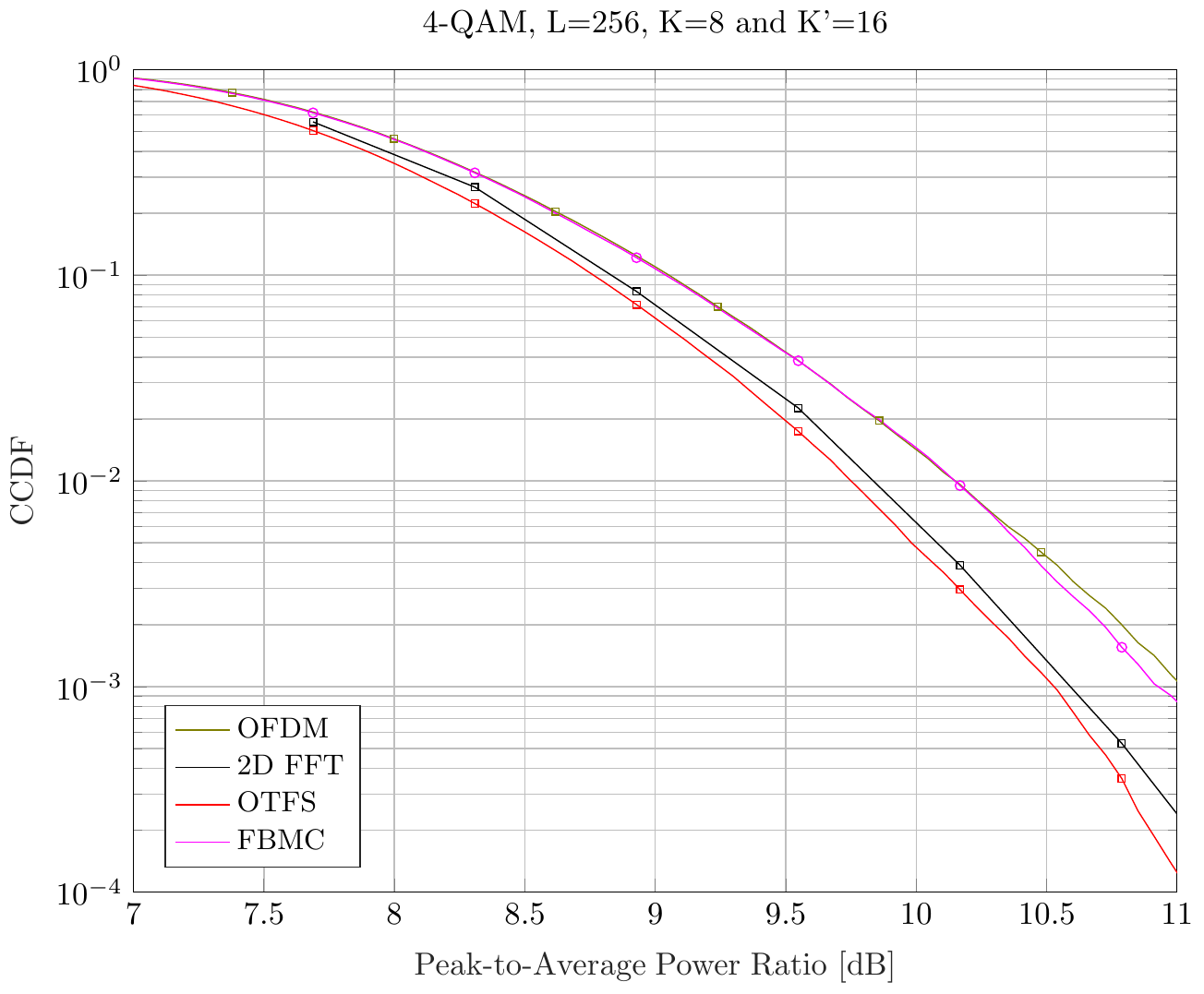}
\caption{PAPR analysis between waveforms.}
\label{PAPR}
\end{figure}

Although OTFS is a multi-carrier system by nature,
its PAPR is much smaller than OFDM and FBMC \cite{surabhi2019peak}. That is because, as shown in~\eqref{OFDM_OTFS}, the PAPR does not depend on the number of subcarriers $L$, but on the size of the block to be transmitted $K$ \cite{hadani2017orthogonal}. In this sense, it is interesting to obtain a good trade-off between complexity and performance, to choose a smaller $K$ with a larger number of subcarriers. For the 2D-FFT FB, the case is similar; however, the PAPR still depends on the number of subcarriers due to the $L$ to $N$ spread needed to obtain filtering. However, such dependence is low, resulting in a PAPR similar to that obtained in OTFS as shown in Figure \ref{PAPR} using L = 128, K' = 16 and a 4-QAM.
\selectlanguage{english}
\selectlanguage{english}
\begin{figure}[!t]
\centering
\includegraphics[width=0.45\textwidth]{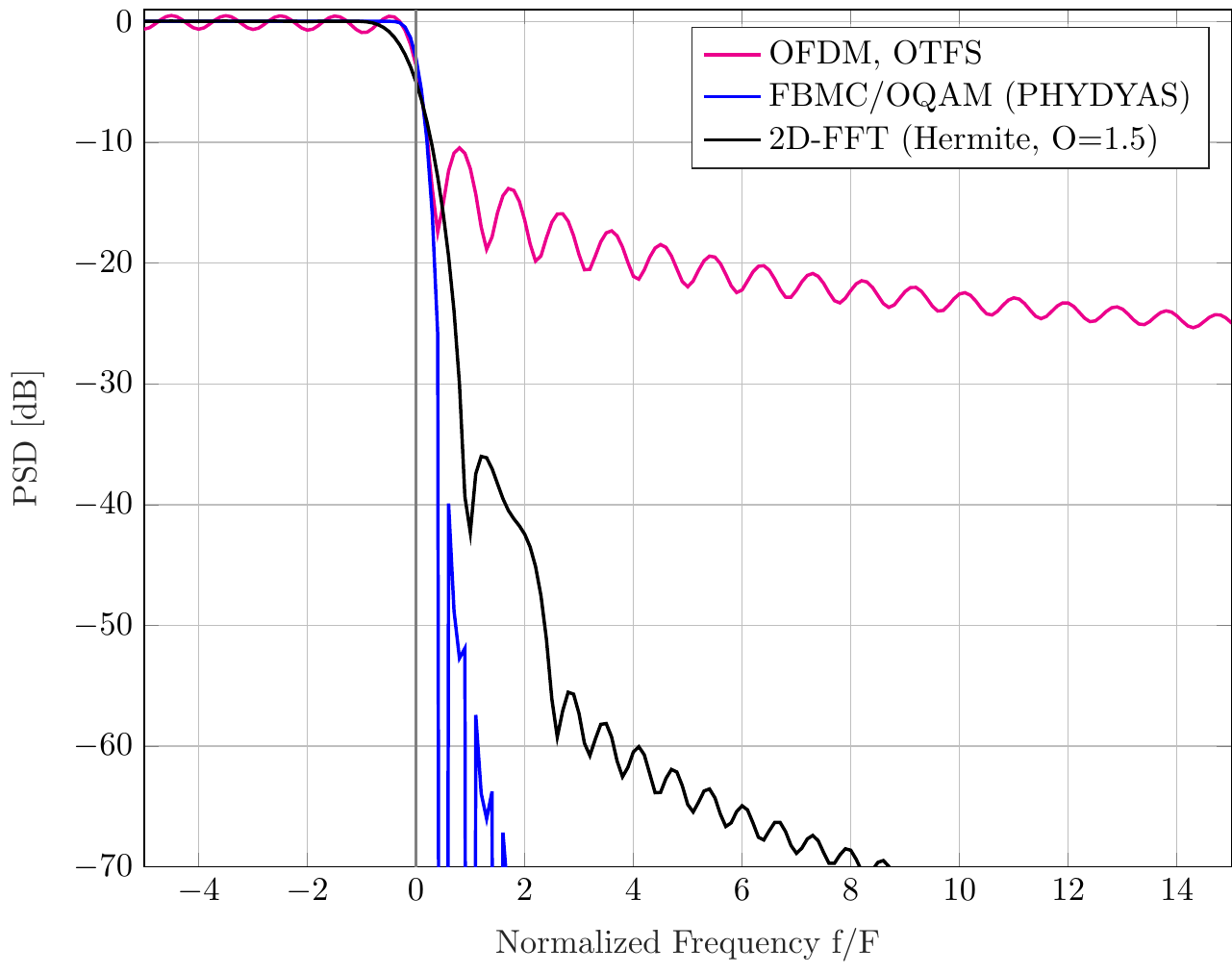}
\caption{Spectral confinement of the analyzed waveforms.}
\label{PSD}
\end{figure}
The inclusion of the filtering process, which improves the spectral confinement as shown in Figure \ref{PSD}, is one of the great advantages of the proposed system. The influence of the side lobes can easily be observed in the frequency domain, as well as the advantage, in OOB terms, of using a filter well located in the frequency. For FBMC/OQAM, we have no restriction of the filter overlapping factor. In this case, using the PHYDYAS filter proposed in \cite{bellanger2010fbmc}, there is a larger overlap factor ($O = 4$), which consequently achieves a better spectral localization. However, even using a filter within the complex orthogonality limit, that is, $O = 1.5$, the OOB emissions of the proposed scheme are comparable with FBMC/OQAM transmissions and much better than those of OFDM and OTFS.

\subsection{Performance in Doubly-Selective Channels}

To analyze the error performance of the proposed system, the transmitted signal is affected by a Rayleigh fading channel using the ITU-T Vehicular A model. Perfect channel estimation is assumed at the receiver. 
For comparison with our proposed system, we also present the performance of the OTFS and DFT precoded filter bank system. The remaining simulation parameters are shown in Table \ref{tabela3}.
\begin{table}[!t]
\centering
\begin{tabular}{|c|c|cc}
\hline
\multicolumn{4}{|c|}{General parameters}                                                                                    \\ \hline
\multicolumn{2}{|c|}{Total subcarriers actives ($L$)}  & \multicolumn{2}{c|}{128}                                                   \\ \hline
\multicolumn{2}{|c|}{Subcarrier spacing ($F$)} & \multicolumn{2}{c|}{15 kHz}                                                \\ \hline
\multicolumn{2}{|c|}{Modulation}               & \multicolumn{2}{c|}{4-QAM and 16-QAM}                                      \\ \hline
\multicolumn{2}{|c|}{Channel Model}            & \multicolumn{2}{c|}{ITU-T Vehicular A}                                     \\ \hline
\multicolumn{2}{|c|}{Velocity   ($V$)}         & \multicolumn{2}{c|}{0, 300 and 400 Km/h}                                   \\ \hline
\multicolumn{2}{|c|}{Carrier frequency}        & \multicolumn{2}{c|}{2.5 GHz}                                               \\ \hline
\multicolumn{4}{|c|}{Specific parameters}                                                                                  \\ \hline
\multicolumn{2}{|c|}{OTFS}                     & \multicolumn{2}{c|}{2D-FFT}                                                \\ \hline
Multi-carrier symbols ($K$)         & 8         & \multicolumn{1}{c|}{Multi-carrier symbols ($K'$)} & \multicolumn{1}{c|}{16}  \\ \hline
Cyclic prefix length          & 8         &  & \multicolumn{1}{c|}{}  \\ \hline
\end{tabular}
\caption{Simulation parameters.}
\label{tabela3}
\end{table}
Note that the FFT used in the proposed system is twice the size ($N = 2L$) of that used in OTFS. Note also that we need to use $N > L$ for the filtering process to be carried out, as explained in Section III.  
However, in practice the systems never operate at a critical sampling rate ($N = L$), that is, the size of the FFT will always be greater than the number of active subcarriers. 
This has many important implications, especially for the channel estimate. In this way, we could choose to use $N = 256$ also for OTFS because in terms of BER results the performance is equivalent.

\begin{figure}[!t]
\centering
\includegraphics[width=0.47\textwidth]{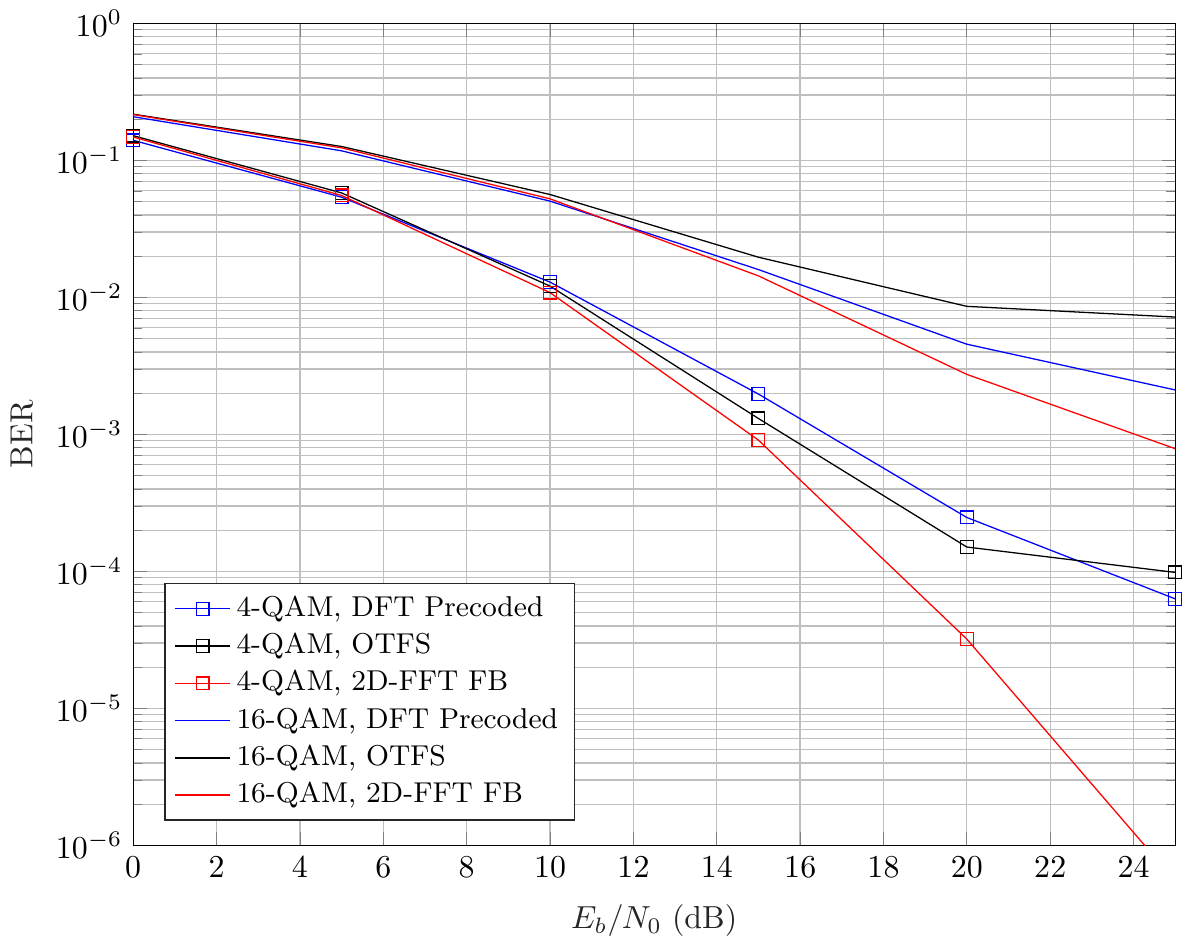}
\caption{Error performance comparison using  frequency domain equalizer (MMSE) for a velocity of 300 km/h.}
\label{BER1}
\end{figure}

\begin{figure}[!b]
\centering
\includegraphics[width=0.47\textwidth]{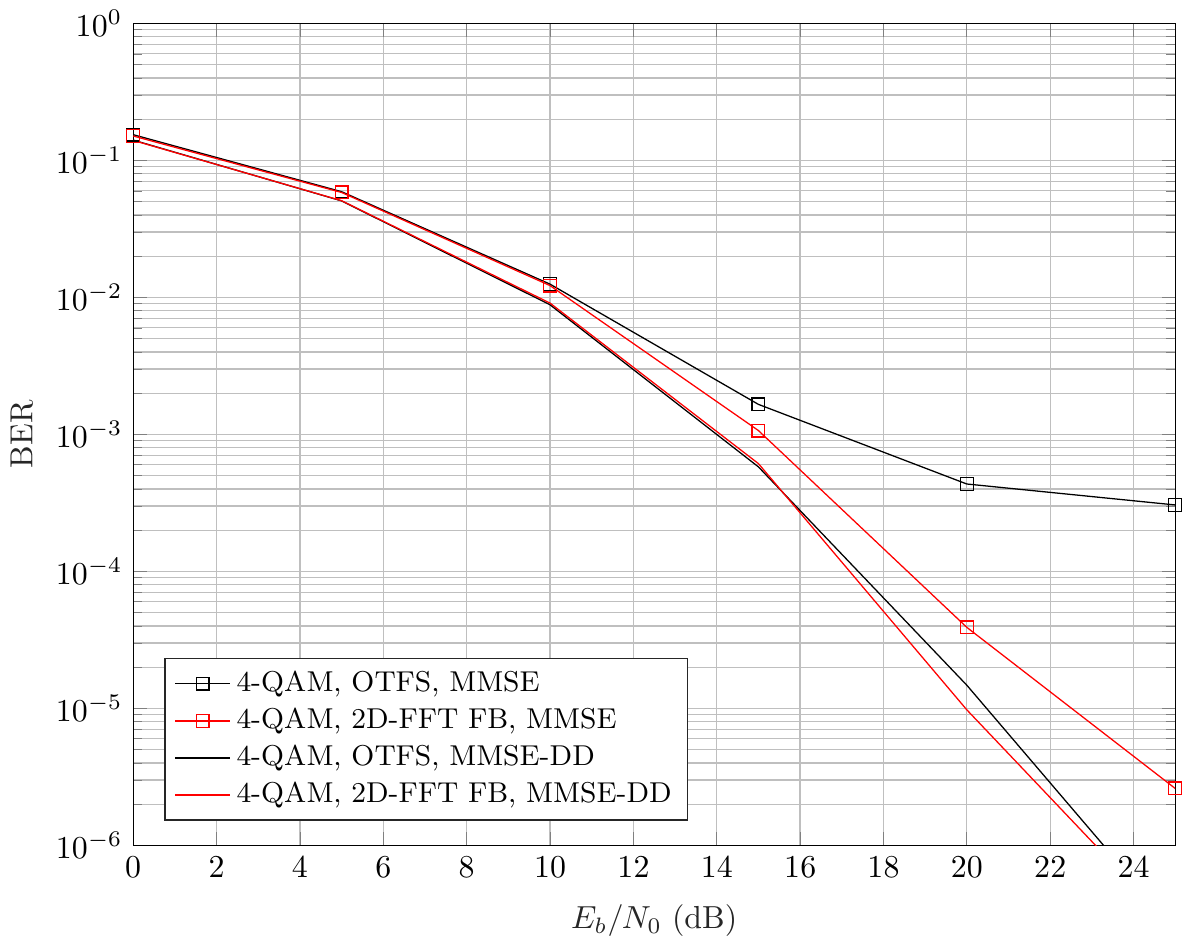}
\caption{Error performance comparison using frequency domain equalizer (MMSE) and delay-Doppler domain equalizer (MMSE-DD) for 4-QAM and velocity of 400 km/h.}
\label{BER2}
\end{figure}

Figure \ref{BER1} shows the uncoded BER for the three systems using a frequency domain equalizer and for a speed of 300 km/h. 
As we can see our system has superior performance for both 4-QAM and 16-QAM modulation.
In addition, the OTFS system features a considerable performance loss when increasing the modulation order bringing an error floor from 25 dB. 
This behavior of the OTFS system is due to the fact that the frequency domain equalizer does not allow the system to extract all channel diversity, which normally occurs when equalization is performed in the delay-Doppler domain. Figure \ref{BER2} presents the MMSE equalizer in both the frequency domain and the delay-Doppler domain (represented by MMSE-DD) for the proposed system compared to the OTFS for 4-QAM and a speed of 400 km/h. 
As we can see, the performance  in the delay Doppler domain of both systems is practically the same. However, this result is obtained at the expense of greater computational complexity. Moreover, at this speed the OTFS system with a frequency domain equalizer has even greater performance loss when compared to the proposed system with the same equalizer.

\begin{figure}[!t]
\centering
\includegraphics[width=0.4\textwidth]{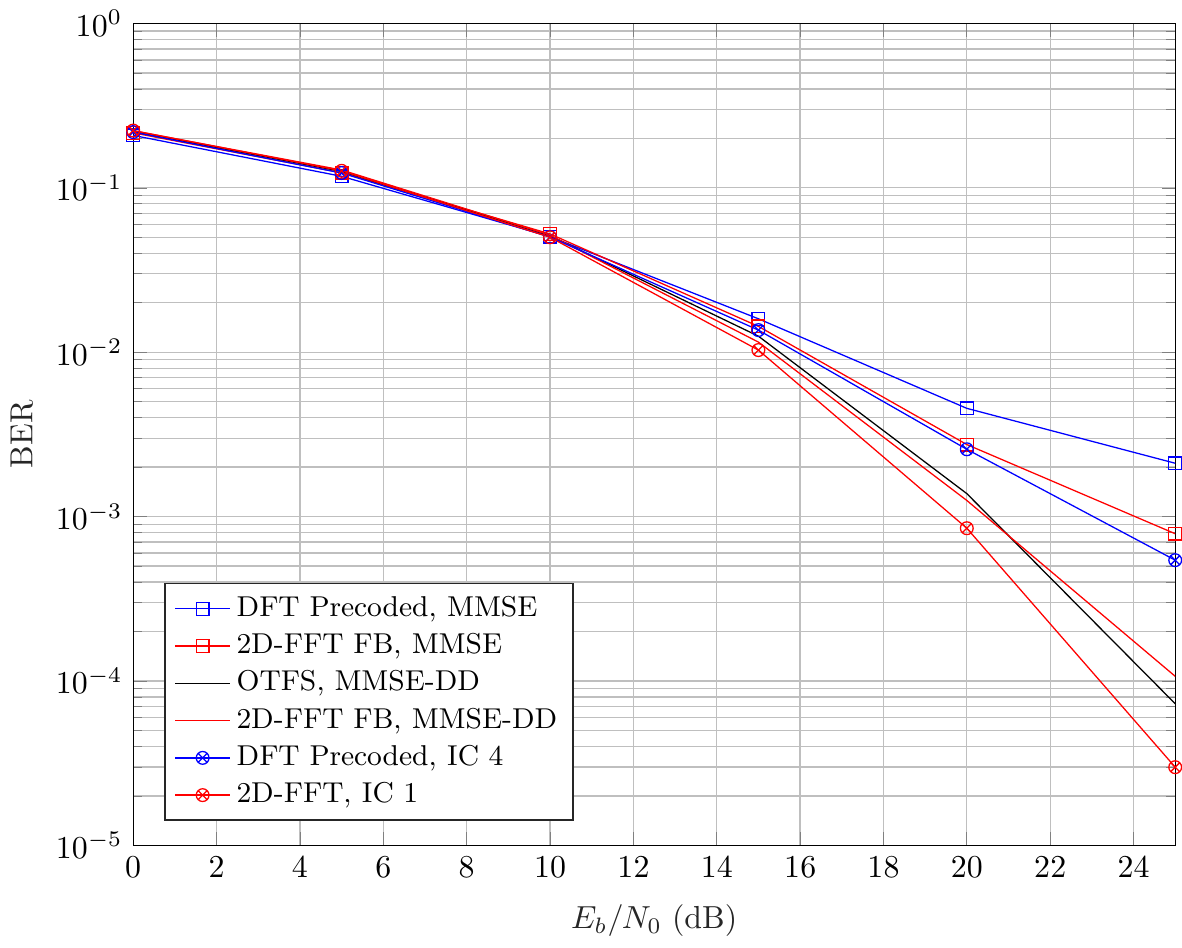}
\caption{Error performance using different receivers for 16-QAM and a velocity of 300 km/h.}
\label{BER3}
\end{figure}

Figure \ref{BER3} presents the results using the proposed IIC receiver for a speed of 300 km/h and a 16-QAM modulation.
Note that the IIC receiver has less computational complexity than the full delay Doppler one applied to OTFS and still has superior performance with just one iteration.
It is worth to note that this hybrid receiver cannot be used with a OTFS transmission, due to the error propagation phenomenon caused by the use of frequency domain equalization (which is the first stage of the proposed IIC equalizer). It is also noteworthy that the IIC receiver does not significantly improve the performance of the DFT precoded system.

\section{Conclusion}
\label{sec:conclusion}

In this work, we present a new precoded filter bank structure based on OTFS modulation, which presents significant advantages in doubly selective channels. The proposed study focuses on the impact of the interference rejection from a channel with large Doppler spread, that is, applications in high mobility scenario. Allied to a PAPR similar to OTFS and a better spectral localization, the proposed system is a good alternative for emerging applications. 
Furthermore, we present a hybrid receiver that uses an equalization stage in the frequency domain combined with an interference canceller in the delay-Doppler domain. 
A clear and concise matrix analysis that allows analyzing the system in BER terms was presented to validate the advantages of the proposed system. Finally, a performance evaluation was performed comparing the proposed scheme with another multi-carrier system. 
This work opens up several possibilities for studies related to filter bank design for multi-carrier transmission systems in a high mobility scenario.
As a suggestion for future work, we can highlight the application in MIMO systems and the proposition of channel estimation techniques that are suitable for our proposed system. 
Another point to highlight is the good performance of the system even when using simple frequency domain equalizers. The study/analysis of this performance as well as the application of other receivers in this scenario are also suggestions for future work.
Finally, due to the restoration of the complex orthogonality which no longer requires the use of OQAM modulation, the change in the waveform design in terms of transmission through the filter bank can also be investigated.

\balance
{
\selectlanguage{english}
\bibliographystyle{./IEEEtran}
\bibliography{./IEEEabrv,./IEEEexample,./IEEEfull}

\begin{thebibliography}{10}
\providecommand{\url}[1]{#1}
\csname url@samestyle\endcsname
\providecommand{\newblock}{\relax}
\providecommand{\bibinfo}[2]{#2}
\providecommand{\BIBentrySTDinterwordspacing}{\spaceskip=0pt\relax}
\providecommand{\BIBentryALTinterwordstretchfactor}{4}
\providecommand{\BIBentryALTinterwordspacing}{\spaceskip=\fontdimen2\font plus
\BIBentryALTinterwordstretchfactor\fontdimen3\font minus
  \fontdimen4\font\relax}
\providecommand{\BIBforeignlanguage}[2]{{%
\expandafter\ifx\csname l@#1\endcsname\relax
\typeout{** WARNING: IEEEtran.bst: No hyphenation pattern has been}%
\typeout{** loaded for the language `#1'. Using the pattern for}%
\typeout{** the default language instead.}%
\else
\language=\csname l@#1\endcsname
\fi
#2}}
\providecommand{\BIBdecl}{\relax}
\BIBdecl

\bibitem{alliance20155g}
N.~Alliance, ``{5G} white paper,'' \emph{Next generation mobile networks, white
  paper}, vol.~1, 2015.

\bibitem{saad2019vision}
W.~Saad, M.~Bennis, and M.~Chen, ``A vision of {6G} wireless systems:
  Applications, trends, technologies, and open research problems,'' \emph{IEEE
  network}, vol.~34, no.~3, pp. 134--142, 2019.

\bibitem{parkvall2017nr}
S.~Parkvall, E.~Dahlman, A.~Furuskar, and M.~Frenne, ``Nr: The new {5G} radio
  access technology,'' \emph{IEEE Communications Standards Magazine}, vol.~1,
  no.~4, pp. 24--30, 2017.

\bibitem{siohan2002analysis}
P.~Siohan, C.~Siclet, and N.~Lacaille, ``Analysis and design of {OFDM/OQAM}
  systems based on filterbank theory,'' \emph{IEEE transactions on signal
  processing}, vol.~50, no.~5, pp. 1170--1183, 2002.

\bibitem{banelli2014modulation}
P.~Banelli, S.~Buzzi, G.~Colavolpe, A.~Modenini, F.~Rusek, and A.~Ugolini,
  ``Modulation formats and waveforms for {5G} networks: Who will be the heir of
  {OFDM}?: An overview of alternative modulation schemes for improved spectral
  efficiency,'' \emph{IEEE Signal Processing Magazine}, vol.~31, no.~6, pp.
  80--93, 2014.

\bibitem{bolcskei2003orthogonal}
H.~B{\"o}lcskei, ``Orthogonal frequency division multiplexing based on offset
  {QAM},'' in \emph{Advances in Gabor analysis}.\hskip 1em plus 0.5em minus
  0.4em\relax Springer, 2003, pp. 321--352.

\bibitem{lele2010alamouti}
C.~L{\'e}l{\'e}, P.~Siohan, and R.~Legouable, ``The alamouti scheme with
  {CDMA-OFDM/OQAM},'' \emph{EURASIP Journal on Advances in Signal Processing},
  vol. 2010, pp. 1--13, 2010.

\bibitem{strohmer2003optimal}
T.~Strohmer and S.~Beaver, ``Optimal {OFDM} design for time-frequency
  dispersive channels,'' \emph{IEEE Transactions on communications}, vol.~51,
  no.~7, pp. 1111--1122, 2003.

\bibitem{boccardi2014five}
F.~Boccardi, R.~W. Heath, A.~Lozano, T.~L. Marzetta, and P.~Popovski, ``Five
  disruptive technology directions for {5G},'' \emph{IEEE communications
  magazine}, vol.~52, no.~2, pp. 74--80, 2014.

\bibitem{farhang2016ofdm}
B.~Farhang-Boroujeny and H.~Moradi, ``{OFDM} inspired waveforms for {5G},''
  \emph{IEEE Communications Surveys \& Tutorials}, vol.~18, no.~4, pp.
  2474--2492, 2016.

\bibitem{hadani2017orthogonal}
R.~Hadani, S.~Rakib, M.~Tsatsanis, A.~Monk, A.~J. Goldsmith, A.~F. Molisch, and
  R.~Calderbank, ``Orthogonal time frequency space modulation,'' in \emph{2017
  IEEE Wireless Communications and Networking Conference (WCNC)}.\hskip 1em
  plus 0.5em minus 0.4em\relax IEEE, 2017, pp. 1--6.

\bibitem{zakaria2012novel}
R.~Zakaria and D.~Le~Ruyet, ``A novel filter-bank multicarrier scheme to
  mitigate the intrinsic interference: Application to {MIMO} systems,''
  \emph{IEEE Transactions on Wireless Communications}, vol.~11, no.~3, pp.
  1112--1123, 2012.

\bibitem{demmer2017block}
D.~Demmer, R.~Gerzaguet, J.-B. Dor{\'e}, D.~Le~Ruyet, and D.~Kt{\'e}nas,
  ``Block-filtered {OFDM}: a novel waveform for future wireless technologies,''
  in \emph{2017 IEEE International Conference on Communications (ICC)}.\hskip
  1em plus 0.5em minus 0.4em\relax IEEE, 2017, pp. 1--6.

\bibitem{nissel2018pruned}
R.~Nissel and M.~Rupp, ``Pruned {DFT-Spread} {FBMC}: Low {PAPR}, low latency,
  high spectral efficiency,'' \emph{IEEE Transactions on Communications},
  vol.~66, no.~10, pp. 4811--4825, 2018.

\bibitem{pereira2020novel}
R.~P. Junior, C.~A. F.~d. Rocha, B.~S. Chang, and D.~Le~Ruyet, ``A novel {DFT}
  precoded filter bank system with iterative equalization,'' \emph{IEEE
  Wireless Communications Letters}, vol.~10, no.~3, pp. 478--482, 2021.

\bibitem{pereira2022generalized}
R.~Pereira, C.~A. da~Rocha, B.~S. Chang, and D.~Le~Ruyet, ``A generalized {DFT}
  precoded filter bank system,'' \emph{IEEE Wireless Communications Letters},
  2022.

\bibitem{murali2018otfs}
K.~Murali and A.~Chockalingam, ``On {OTFS} modulation for high-doppler fading
  channels,'' in \emph{2018 Information Theory and Applications Workshop
  (ITA)}.\hskip 1em plus 0.5em minus 0.4em\relax IEEE, 2018, pp. 1--10.

\bibitem{hadani2018otfs}
R.~Hadani and A.~Monk, ``{OTFS}: A new generation of modulation addressing the
  challenges of {5G},'' \emph{arXiv preprint arXiv:1802.02623}, 2018.

\bibitem{junior2021iterative}
R.~P. Junior, C.~A.~F. da~Rocha, B.~S. Chang, and D.~Le~Ruyet, ``Iterative
  interference cancellation for the {DFT} precoded filter bank system,'' in
  \emph{2021 17th International Symposium on Wireless Communication Systems
  (ISWCS)}.\hskip 1em plus 0.5em minus 0.4em\relax IEEE, 2021, pp. 1--6.

\bibitem{zemen2017low}
T.~Zemen, M.~Hofer, and D.~Loeschenbrand, ``Low-complexity equalization for
  orthogonal time and frequency signaling ({OTFS}),'' \emph{arXiv preprint
  arXiv:1710.09916}, 2017.

\bibitem{raviteja2018interference}
P.~Raviteja, K.~T. Phan, Y.~Hong, and E.~Viterbo, ``Interference cancellation
  and iterative detection for orthogonal time frequency space modulation,''
  \emph{IEEE Transactions on Wireless Communications}, vol.~17, no.~10, pp.
  6501--6515, 2018.

\bibitem{tiwari2019low}
S.~Tiwari, S.~S. Das, and V.~Rangamgari, ``Low complexity {LMMSE} receiver for
  {OTFS},'' \emph{IEEE communications letters}, vol.~23, no.~12, pp.
  2205--2209, 2019.

\bibitem{haas1997time}
R.~Haas and J.-C. Belfiore, ``A time-frequency well-localized pulse for
  multiple carrier transmission,'' \emph{Wireless personal communications},
  vol.~5, no.~1, pp. 1--18, 1997.

\bibitem{surabhi2019peak}
G.~Surabhi, R.~M. Augustine, and A.~Chockalingam, ``Peak-to-average power ratio
  of {OTFS} modulation,'' \emph{IEEE Communications Letters}, vol.~23, no.~6,
  pp. 999--1002, 2019.

\bibitem{bellanger2010fbmc}
M.~Bellanger, D.~Le~Ruyet, D.~Roviras, M.~Terr{\'e}, J.~Nossek, L.~Baltar,
  Q.~Bai, D.~Waldhauser, M.~Renfors, T.~Ihalainen \emph{et~al.}, ``{FBMC}
  physical layer: a primer,'' \emph{PHYDYAS, January}, vol.~25, no.~4, pp.
  7--10, 2010.

\end{thebibliography}
}

\end{document}